\documentclass[prd,preprint,tightenlines,superscriptaddress,floatfix,showpacs,
preprintnumbers,nofootinbib,eqsecnum]{revtex4}

 \usepackage[dvips,final]{graphicx}
     \usepackage{epsfig}
      \usepackage{bm}% bold math
\begin{document}

\thispagestyle{empty} \preprint{\hbox{}} \vspace*{-10mm}

\title{Spectator induced electromagnetic effect on directed flow \\
in heavy ion collisions}

\author{A.~Rybicki}
\email{andrzej.rybicki@ifj.edu.pl}

\affiliation{H.Niewodnicza\'nski Institute of Nuclear Physics, Polish Academy of Sciences, Radzikowskiego 152, 31-342 Krak\'ow, Poland}

\author{A.~Szczurek}
\email{antoni.szczurek@ifj.edu.pl}

\affiliation{H.Niewodnicza\'nski Institute of Nuclear Physics, Polish Academy of Sciences, Radzikowskiego 152, 31-342 Krak\'ow, Poland}
\affiliation{University of Rzesz\'ow, Rejtana 16, 35-959 Rzesz\'ow, 
Poland}

\date{\today}

\begin{abstract}
 %Abstract. 

We estimate the electromagnetic effect of the spectator charge on
azimuthal anisotropies observed in heavy ion collisions. For peripheral
Pb+Pb reactions at the top energy of the CERN Super Proton Synchrotron, 
$\sqrt{s_{NN}}=17.3$ GeV, we predict this effect to bring very large 
distortions to the observed directed flow, $v_1$, of positive and
negative pions emitted close to beam rapidity. The overall magnitude of 
this effect is comparable to values of $v_1$ reported by the WA98 experiment.
We argue that also at lower rapidities, the spectator induced
electromagnetic effect may result in the splitting of values of $v_1$ 
observed for positive and negative pions. 
Such a splitting is visible in the data reported by the STAR
Collaboration from the RHIC Beam Energy Scan.

Both effects are sensitive to the space-time scenario assumed for 
pion emission. Therefore, they bring new information on the collision dynamics.

\end{abstract}

\pacs{25.75.-q,25.75Dw,25.75.Ld}

%\pacs{12.38.Bx, %(Quantum chromodynamics: Perturbative calculations) \\
%                                24.85.+p, %(Quarks, gluons, and QCD in nuclear reactions)\\
 %                25.20.Lj, %(Nuclear reactions: specific reactions: Photoproduction reactions)\\
  %               25.75.Dw. %(Relativistic heavy-ion collisions: Particle and resonance production)
        %}

\maketitle

%----------------------------
\section{Introduction}
\label{secone}
%----------------------------

Azimuthal correlations between particles and the reaction plane, 
observed in non-central collisions, constitute one of the 
main subjects of heavy ion collision studies. The primary 
nucleon-nucleon collisions being believed to be insensitive to the 
reaction plane, the latter phenomena are considered to be the source of 
information on {\em collective effects} present in the heavy ion 
reaction. A rich phenomenology~\cite{Snellings1999,Csernai2012,Luzum2010,Retinskaya2012,florkowski} has therefore been 
developed. Among others, the success of the hydrodynamical approach 
~\cite{bozek} in describing these phenomena is to be quoted here.

The azimuthal correlations are usually quantified in terms of the Fourier coefficients of the azimuthal distribution of the outgoing particles w.r.t.~the reaction plane. These are defined by~\cite{poskanzerprc}:

\begin{equation}
 v_n \equiv \langle \cos [n (\phi - \Psi_r)] \rangle   \; \; ,
\end{equation}
 where $\phi$ denotes the azimuthal angle of the emitted particle, while $\Psi_r$ is the orientation of the reaction plane defined by the impact parameter vector $\vec{b}$. Specifically, the first-order coefficient 

\begin{equation}
 v_1 \equiv \langle \cos (\phi - \Psi_r) \rangle
\end{equation}
 reflects the {\em sidewards collective motion} of the particles and is known as directed flow. A very sizeable amount of experimental data on directed flow is at present available from the low energy regime up to the LHC~\cite{Andronic2001,Barrette2000,Barrette1998,na49prc,Adams2005,Back2005,Abelev2008,Adamczyk2011,Selyuzhenkov2011}. It is known that $v_1$ depends on collision centrality, particle type, transverse momentum and rapidity. For instance, the rapidity dependence of pion directed flow at SPS energies is known to consist of a smooth passage through zero at mid-rapidity~\cite{na49prc}, while large values of $v_1$, of the order of 0.2-0.25, are reported
 % for positive pions
 in the vicinity of target rapidity, in semi-central collisions~\cite{wa98}.

In the present paper we point at the possible {\em electromagnetic component} of directed flow.
 % of charged pions. 
 We estimate the size of the spectator-induced electromagnetic interaction on the collective motion of positive and negative pions. The spectator-induced electromagnetic effect on charged pion ratios $\left(\frac{\pi^+}{\pi^-}\right)$ was discussed in our earlier paper~\cite{twospec}. This effect is in fact observed in experimental data~\cite{rybickizako}, and results in very large distortions in ratios of charged pions observed at high rapidities and low transverse momenta. Measurements of directed flow offer a new way of spotting the influence of the spectator charge on the emission of final state particles. They also provide a new source of information
% about 
 on the space-time evolution of the non-perturbative process of pion
 production\footnote{The 
cited paper~\cite{twospec} also contains a description of various works made in the past on electromagnetic effects in low and high energy nuclear collisions. For more details on this subject,
 see~\cite{Bartke09,Yagoda52,Friedlander62,Beneson79,Sullivan82,Karnaukhov06,Ayala97,Ayala99,Ahle98,Lacasse96,Xu96,Libbrecht79,Cugnon81,Gyulassy81,Bonasera87,Li95,Barz98,Osada96}.}.

Our present analysis will be focussed on peripheral Pb+Pb collisions occurring at the top energy available to SPS heavy ion experiments ($\sqrt{s_{NN}}=17.3$~GeV). We choose peripheral collisions because they are characterized by the largest spectator charge and also large values of directed flow. We focus on the SPS energy range by virtue of the abundance of experimental data on the rapidity dependence of $v_1$, which also gives the possibility of
 %a tentative
 comparison with our results as explained in Section~\ref{results}.
 We note that the energy range of the SPS partially overlaps with that
 of the RHIC Beam Energy Scan, addressed in short in Section~\ref{bes}.

%\newpage

This paper is organized as follows. Section~\ref{model} contains the description of our Monte Carlo model. In Section~\ref{results} we present the results of our simulation, together with the comparison to experimental data. The discussion of our results is made in Section~\ref{discussion}. 
In Section~\ref{bes}, we comment
on the experimental data from the RHIC Beam Energy Scan.
Our conclusions are presented in Section~\ref{conclusions}.

%----------------------------
\section{The model}
\label{model}
%----------------------------

A detailed description of our model can be found in~\cite{twospec}. Only the aspects that are relevant for the present analysis will be listed below.
Our aim is to obtain a realistic estimate of the influence of the spectator-induced electromagnetic interaction on the directed flow of pions. On the other hand, we wish to avoid the detailed discussion of the complex and poorly known mechanism of soft particle production. Therefore we decide on a maximally simplified approach:

%; this is illustrated in Fig.~X1.

%%%%%%%%%%%%%%%%%%%%%%%%%%%%%%%%%%%%%%%%%%%%%%%%%%%%%%%%%%%%%
\begin{figure}[t]             
%\begin{minipage}[t]{0.46\textwidth}
\centering
\includegraphics[width=0.75\textwidth]{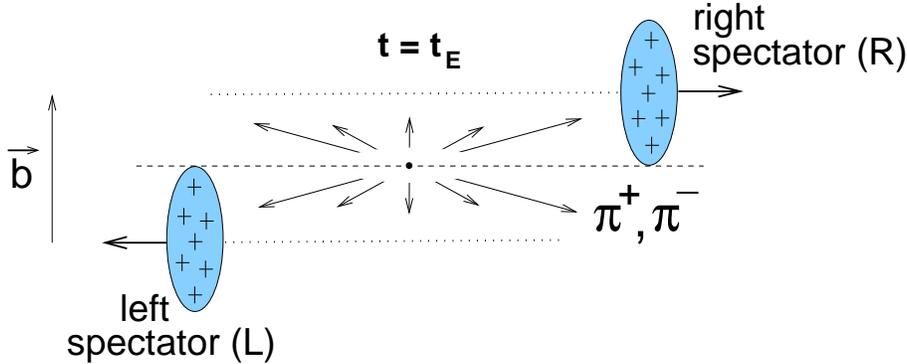}
%\end{minipage}
   \caption{(Color online) Our model of the peripheral Pb+Pb collision. The two spectator systems are denoted left (L) and right (R), respectively.}
 \label{fig:model}
%           }
\end{figure}
%%%%%%%%%%%%%%%%%%%%%%%%%%%%%%%%%%%%%%%%%%%%%%%%%%%%%%%%%%%%%%%%%%

\begin{enumerate}
\item[(a)] 
We assume, as an example, a peripheral Pb+Pb collision involving 
$n_{part}=60$ participating nucleons at $\sqrt{s_{NN}}=17.3$~GeV. This 
is shown in Fig.~\ref{fig:model}. 
The two spectator systems are modelled as two homogeneous, Lorentz-contracted spheres. The reaction plane is defined by the collision axis (dashed line in the figure) and the impact parameter vector~$\vec{b}$.
% (Fig.~1x). 
%In the rest frame of each sphere, its density is $\rho=0.17$/fm$^\mathrm{3}$. The charge of each sphere is 70 elementary units. This situation is shown in Fig.~X1. We note that the reaction plane is defined by the collision axis (dashed line in the figure) and the impact parameter vector~{\bf b}.

\item[(b)] 
Charged pions are assumed to be emitted from a single point in position space, that is, the original interaction point. The time of pion emission $t_E$ is a free parameter of our model; this parameter sets the initial conditions for the electromagnetic interaction. For peripheral Pb+Pb collisions studied here, the initial spectra of charged pions are assumed to be similar to those in nucleon-nucleon collisions.

\item[(c)] 
Charged pions are then numerically traced in the electromagnetic field of the spectator charges
 until they reach the distance of 10,000 fm away from the interaction point and from each of the two spectator systems. 
%This results in a modification of their momentum vector. 
The fragmentation of the spectator systems
%, as well as other second-order effects, are 
is
neglected.
We do not consider the effects of participant charge, strong final state interactions, etc.

%and from each of the spectator system - IN ITS REST FRAME?

% of our model and will be discussed in Section~\ref{results}.

\end{enumerate}

%\newpage

Several clarifications should be added to the above.

\begin{enumerate}
\item[(a)] 
The simplified geometry of the peripheral collision presented in Fig.~\ref{fig:model} was determined on the basis of a dedicated study, discussed in detail in~\cite{twospec}. A geometrical Monte Carlo simulation served this purpose. This used proper nuclear density profiles~\cite{mizutori} and assumed the elementary nucleon-nucleon cross-section equal to 31.4~mb, in good agreement with experimental p+p data at this collision energy~\cite{pp}. The geometrical impact parameter corresponding to 60 participating nucleons was found to be $b_{geom}=10.61$~fm. Additionally, the center of gravity of each of the two spectator systems was found to be displaced by 
%0.76~fm
$\Delta b=0.76$~fm 
relative to that of the original nucleus. 
The average spectator charge was $Q=70$ elementary units.\\
Considering the exact shape of the spectator system as unimportant for our subsequent studies, and for the sake of clarity, we modelled the two spectator systems as homogeneously charged spheres. In the rest frame of each sphere, its density was the standard nuclear density $\rho=0.17$/fm$^\mathrm{3}$. The center of each sphere was additionally displaced by 0.76 fm in order to match the center of gravity of the spectator system.
Thus our effective impact parameter (distance of closest approach between the two spheres' centers) was $b=b_{geom}+2\Delta b=12.13$~fm. 

\item[(b)] 
The simplification of initial conditions for pion emission (reduction of the emission zone to a single point in space and time, as discussed above) gives a convenient way to estimate the sensitivity of the electromagnetic effect to the basic characteristics of pion production (the pion formation time, the distance between the pion formation zone and the two spectator systems, etc). 
The initial momentum spectra of pions (before the action of the electromagnetic field) are assumed to obey wounded nucleon scaling~\cite{wnm} and to be similar to underlying nucleon-nucleon events. As such, they are described by an analytical parametrization of average pion $\left(\frac{\pi^++\pi^-}{2}\right)$ distributions in p+p collisions, recorded by the NA49 experiment at the SPS~\cite{pp}. More details on this parametrization are given in~\cite{twospec}. For simplicity, 
 %and in view of the approximate isospin symmetry of the Pb nucleus, 
distributions of positive and negative pions are assumed to be identical. The experimental data cited above are expressed in terms of the Feynman variable\footnote{All the kinematical variables addressed in this paper will be defined in the c.m. system of the collision.} $x_F=2p_L/\sqrt{s}$ and of transverse momentum $p_T$, and cover the region from $x_F=0$ to $0.85$ and from $p_T=0$ to $2.1$~GeV/c. Only a small extrapolation towards higher $x_F$ needs to be applied in the present analysis. The uncertainty of this extrapolation has little or no effect on the results presented in Section~\ref{results}. It should be underlined that full azimuthal symmetry is assumed for the initial emission of pions.

\item[(c)]
Charged pions, with their initial momentum vector defined above in point (b), and weighted by their initial distribution, are subjected to the electromagnetic field of the two spectator systems. The pion trajectory $\vec{r}_{\pi}$ is given by the classical relativistic equation of motion:

\begin{equation}
\frac{d \vec{r}_{\pi}}{d t} = \vec{v}_{\pi}(\vec{r},t) = 
\frac{\vec{p}_\pi \; c^2}{\sqrt{p^2_\pi \; c^2 + m^2_\pi \; c^4 }} \; ,
\label{velocity}
\end{equation}

where $m_{\pi}$ is the pion mass, and the pion momentum is defined by the Lorentz force acting on the pion:

\begin{equation}
\frac{d \vec{p}_{\pi}}{dt} = \vec{F}_{\pi}(\vec{r},t) =
 q_{\pi} \left( \vec{E}(\vec{r},t) 
+ {\vec{v}_{\pi}(\vec{r},t)} \times \vec{B}(\vec{r},t) \right) \; .
\label{Lorentz_force}
\end{equation}

Here, $q_{\pi}$ is the pion charge, while $\vec{E}(\vec{r},t)$ and $\vec{B}(\vec{r},t)$ are standard superpositions of fields from the two spectator systems. Our equations take account of relativistic effects, including retardation. Technically, the propagation of the pion is performed numerically by means of an iterative procedure made in small steps in time, with variable step size. The iteration proceeds until the pion is 10,000 fm away from the interaction point and from each of the two spectator systems in their respective rest frames. Negative pions which do not escape from the potential well induced by the spectator system are rejected and do not enter
 into
 our final state distributions. 

\end{enumerate}

%----------------------------
\section{Results}
\label{results}
%----------------------------

In this Section we present the results of our analysis. These will be displayed as a function of the scaled rapidity $y/y_\mathrm{beam}$, where $y$ and $y_\mathrm{beam}$ are the rapidity of the pion and of the incoming nucleus in the collision c.m. system. We decide on this particular variable in order to simplify the interpretation of our results, and for an easier possible comparison with other collision energies (see also the discussion made in~\cite{na49prc,star-Pandit}).

%\newpage

%-------------
\subsection{Electromagnetic effect from one and two spectators}
\label{onespec}
%-------------

%%%%%%%%%%%%%%%%%%%%%%%%%%%%%%%%%%%%%%%%%%%%%%%%%%%%%%%%%%%%%
\begin{figure}[t]             
%\begin{minipage}[t]{0.46\textwidth}
\centering
\includegraphics[width=0.95\textwidth]{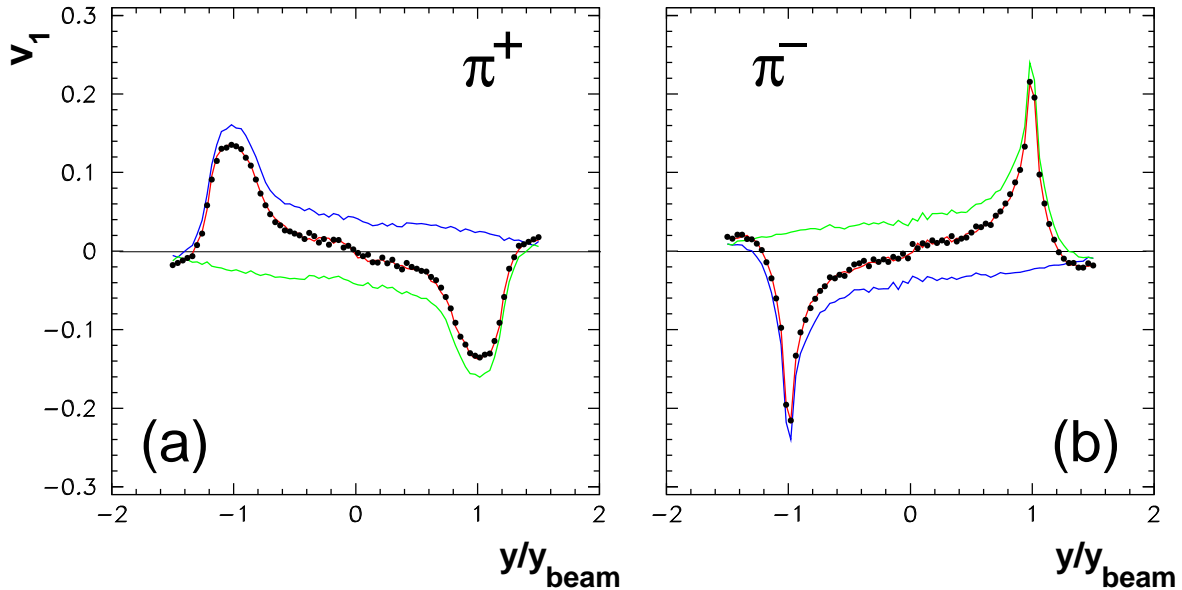}
%\end{minipage}
   \caption{(Color online) Spectator-induced electromagnetic effect on directed flow of (a) positive and (b) negative pions, in peripheral Pb+Pb collisions at $\sqrt{s_{NN}}=17.3$~GeV. The green solid curve shows the directed flow induced electromagnetically by the right (R) spectator. The blue solid curve shows the directed flow induced electromagnetically by the left (L) spectator. Black dots show the result of the addition of these two curves. The red solid curve displays the result of the simulation including both spectators. Note: all the simulations assume $t_E=0$~fm/c; the blue solid curve is obtained by reflection of the green solid curve about midrapidity.}
%$y/y_\mathrm{beam}=0$.}
 \label{fig:onespec}
%           }
\end{figure}
%%%%%%%%%%%%%%%%%%%%%%%%%%%%%%%%%%%%%%%%%%%%%%%%%%%%%%%%%%%%%%%%%%

We start by discussing the basic features of the spectator-induced electromagnetic effect on the directed flow of pions. In particular, this will include the role of each of the two spectator systems in the overall effect. Fig.~\ref{fig:onespec} shows the directed flow $v_1$ of $\pi^+$ and $\pi^-$, induced by their electromagnetic interaction with the spectator charges. The value of $v_1$ is integrated over the transverse momentum of the pion, from $p_T=0$ to 1 GeV/c.
 All the simulations presented in the figure assume the simplest situation where the pion emission time $t_E$ is equal to zero (immediate pion creation). 

It should be underlined that as our simulation contains no initial azimuthal anisotropy (Sect.~\ref{model}), any non-zero value of $v_1$ apparent in Fig.~\ref{fig:onespec} is indeed {\em solely due to the electromagnetic interaction with the spectator systems}. Therefore, we will refer to it as ``electromagnetically-induced directed flow'' in order to differentiate it from the ``standard'' flow phenomena which are caused by the strong interaction.
In this context, several remarks are in order. 

\begin{enumerate}
\item
 As it is clearly apparent from the figure, the electromagnetic field induced by the spectator charges exerts a strong influence on the directed flow of charged pions. 
%Account taken of the fact that our simulation contains no initial azimuthal anisotropy (Sect.~\ref{model}), any non-zero value of $v_1$ is induced {\em solely by the electromagnetic interaction}. Thus, the latter interaction 
 %In peripheral Pb+Pb collisions discussed here,
It can be held responsible for large values of $v_1$, exceeding 0.2 (20\%) for negative pions close to beam rapidity.
\item
 The directed flow induced electromagnetically by the two spectators (red solid curve, which goes through the black dots in the figure) exhibits a characteristic structure as a function of rapidity. Qualitatively, this structure 
 %can be said to recall 
 recalls
 the overall behaviour of pion directed flow as it is reported from experimental data (see e.g.~\cite{wa98,na49prc}). The directed flow displays a smooth transition through zero at $y/y_\mathrm{beam}=0$, with large values of flow (anti-flow) at large negative (positive) rapidity. 
\item
The sign of electromagnetically-induced directed flow for negative pions is opposite with respect to positive pions. For absolute values of $v_1$, differences in the shape of the two curves are seen close to beam rapidity.
\item
What specifically follows from the above is that the spectator-induced electromagnetic interaction may result in the {\em splitting} of values of $v_1$ observed for positive and negative pions. This splitting exhibits a strong rapidity-dependence, and reach very large values close to beam rapidity.
\item
The directed flow which is induced electromagnetically by each of the two spectators {\em separately} is also presented in the figure (the two spectator systems are denoted ``left'' and ``right'' as in Fig.~\ref{fig:model}). The contribution of a single spectator does not remain confined to its hemisphere of the collision. On the contrary, it extends into the opposite hemisphere; thus the ``right'' spectator moving at positive rapidity exerts its influence on the pion $v_1$ down to and beyond $y/y_\mathrm{beam}=-1$, and vice-versa for the ``left'' spectator. 

%This is somewhat surprising in view of our precedent studies which indicated that the spectator-induced electromagnetic effect on $\pi^+/\pi^-$ ratios was small, if any, at $y=0$~\cite{twospec}.

\item
The value of $v_1$ obtained as the result of the electromagnetic interaction with both spectators (red solid curve, which goes through black dots in the figure) appears equal to the result of direct addition of the two single-spectator curves (displayed by the black dots).

%\item
%Finally, an interesting feature of the electromagnetically-induced directed flow is apparent from the figure: the value of $v_1$ resulting from the influence of both spectators (red solid curve) appears equal to the result of a direct addition of the two single-spectator curves (this result is given by the black dots in the figure).

\end{enumerate}

Thus, as it it becomes evident from the above, the spectator-induced electromagnetic interaction appears strong enough to result in sizeable effects, and to play an important role in flow phenomena involving charged pions in heavy ion collisions.
This imposes the necessity of more detailed studies which will be presented below.

%-------------
\subsection{Dependence on initial conditions}
\label{initial}
%-------------

%%%%%%%%%%%%%%%%%%%%%%%%%%%%%%%%%%%%%%%%%%%%%%%%%%%%%%%%%%%%%
\begin{figure}[t]             
%\begin{minipage}[t]{0.46\textwidth}
\centering
\includegraphics[width=0.95\textwidth]{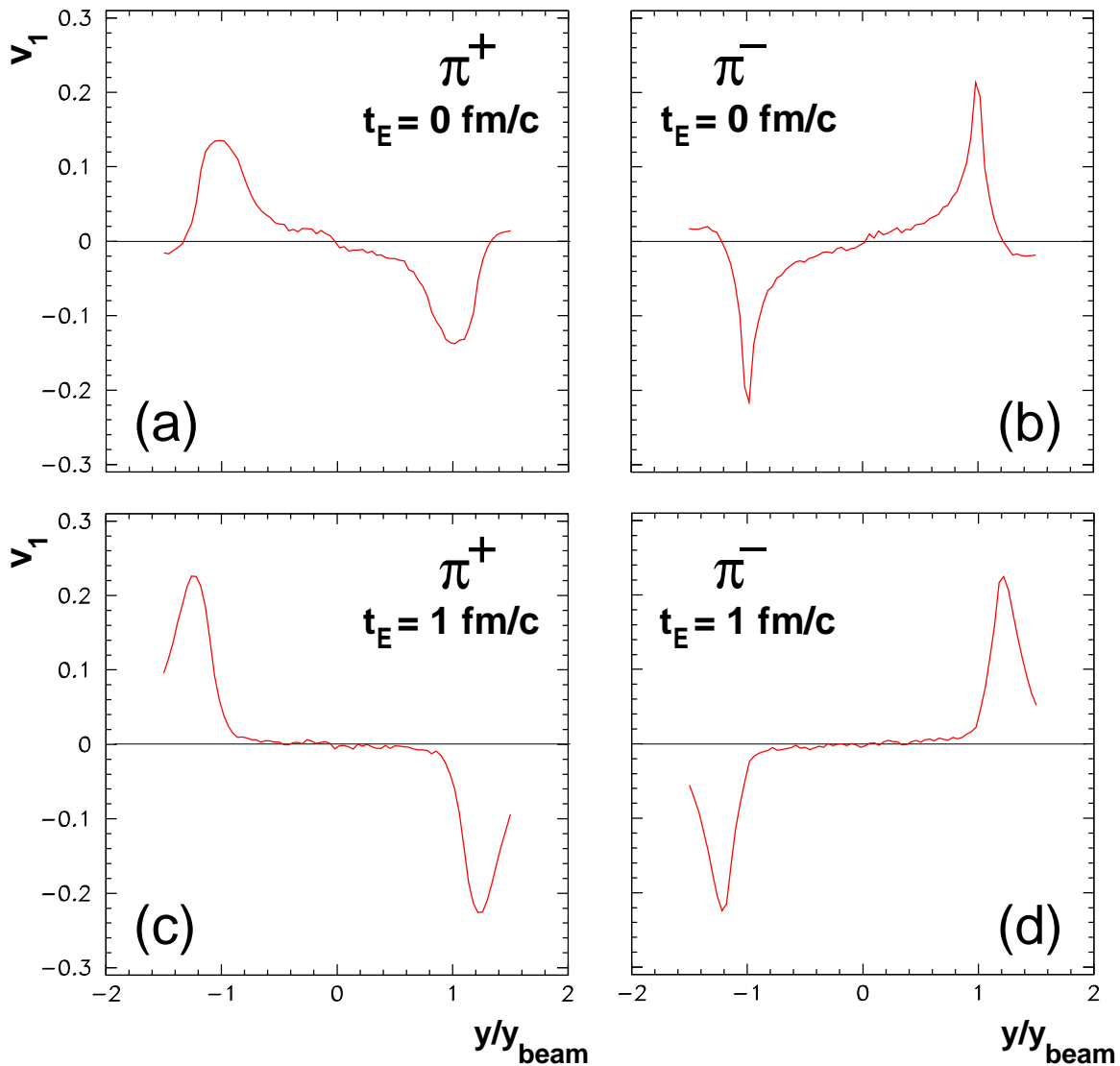}
%\end{minipage}
   \caption{(Color online) Spectator-induced electromagnetic effect on directed flow of positive and negative pions, in peripheral Pb+Pb collisions at $\sqrt{s_{NN}}=17.3$~GeV. The top and bottom panels correspond to different values of the pion emission time $t_E$.}
 \label{fig:initial}
%           }
\end{figure}
%%%%%%%%%%%%%%%%%%%%%%%%%%%%%%%%%%%%%%%%%%%%%%%%%%%%%%%%%%%%%%%%%%

%\newpage

The central issue of this paper is the sensitivity of the spectator-induced electromagnetic effect 
 to {\em the space-time evolution of 
 the collision}. 
This will be addressed in the present Section.

Fig.~\ref{fig:initial} shows the rapidity-dependence of the
electromagnetically-induced directed flow of positive and negative
pions, obtained assuming two different values of the pion emission time:
$t_E=0$ and $1$~fm/c. The values of $v_1$ are obtained by integration over $p_T$ of the pion, with integration limits defined as in Section~\ref{onespec} above. The curves in panels (a) and (b) of the figure are the same as the respective curves in Fig.~\ref{fig:onespec}. The following remarks are to be made.

\begin{enumerate}
\item
The electromagnetically-induced directed flow displays a clear dependence on the pion emission time.
 In the region of beam rapidity, the position of the minimum (maximum) in the valley (peak) in pion $v_1$ shifts from $y/y_\mathrm{beam}=1$ at $t_E=0$~fm/c to above $y/y_\mathrm{beam}=1.2$ at $t_E=1$~fm/c. 
 At the collision energy discussed here,
 this shift in $y/y_\mathrm{beam}$ corresponds to a shift of about 0.6 units in rapidity. The actual shape of the valley (peak) also exhibits sensitivity to $t_E$.
 \item
 Also in the region of lower rapidities, and down to midrapidity, the change in $t_E$ results in changes of $v_1$. For instance, at $y/y_\mathrm{beam}=0.5$ the absolute values of $v_1$ go from about 2.4\% at $t_E=0$~fm/c down to about 0.5\% at $t_E=1$~fm/c.
 \item
What follows from the above is that the electromagnetically-induced splitting of $v_1$ for $\pi^+$ and $\pi^-$, mentioned in Section~\ref{onespec},
 will also exhibit  
sensitivity to $t_E$.
\end{enumerate}

These observations demonstrate that the spectator-induced electromagnetic effect 
%on directed flow
 {is indeed sensitive to the initial conditions, set for the electromagnetic interaction} by the emission time $t_E$.
Changes in the pion emission time, or equivalently in the position of the pion formation zone with respect to the two spectator systems, will result in changes of $v_1$ which should be observable in experiment. In other terms, the electromagnetic effect on directed flow {\em is indeed sensitive to the space-time evolution of pion production}, and provides new information on the dynamics of the collision.

%-------------
\subsection{Dependence on transverse momentum}
\label{ptdep}
%-------------

%%%%%%%%%%%%%%%%%%%%%%%%%%%%%%%%%%%%%%%%%%%%%%%%%%%%%%%%%%%%%
\begin{figure}[t]             
%\begin{minipage}[t]{0.46\textwidth}
\centering
\includegraphics[width=0.95\textwidth]{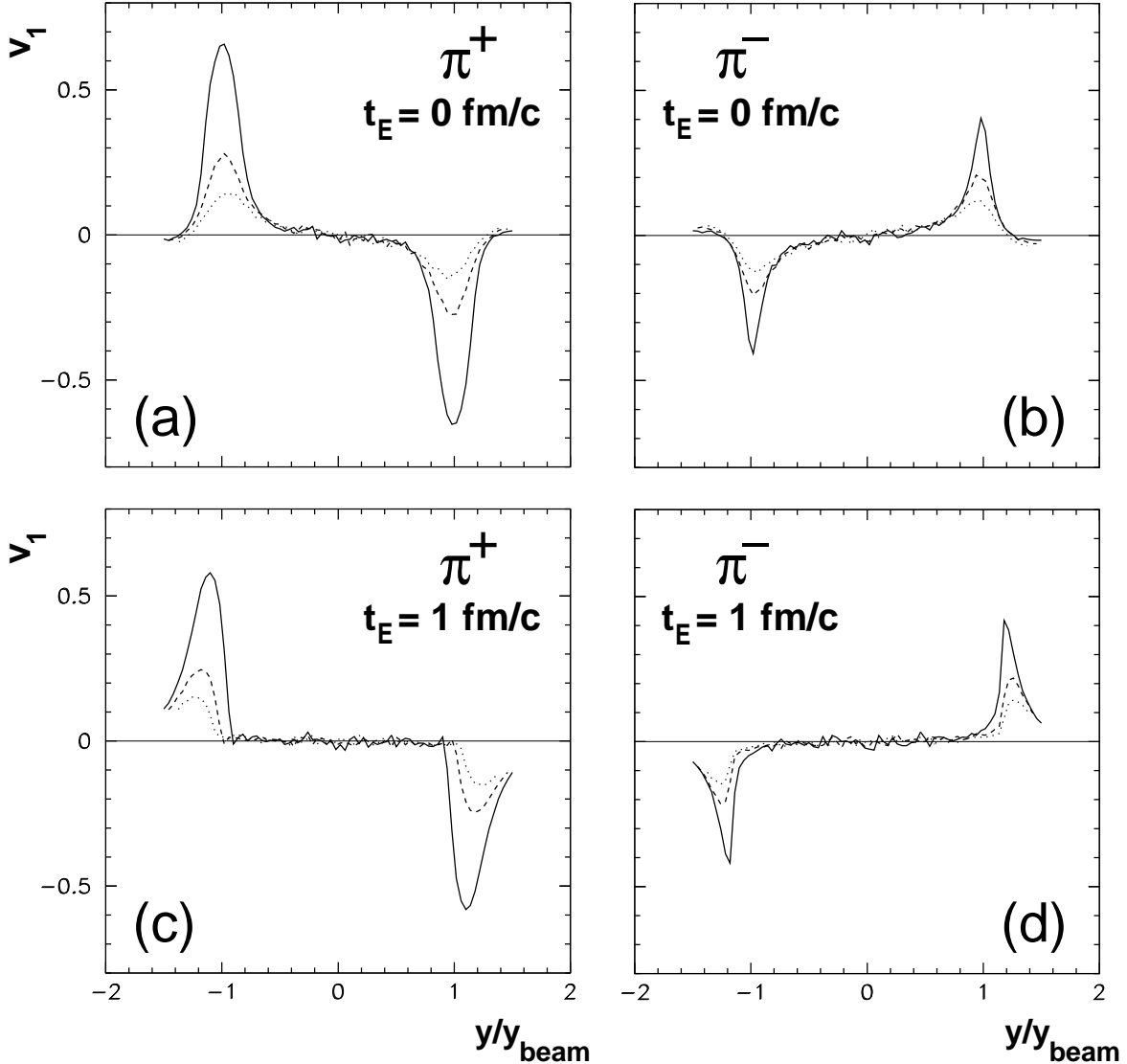}
%\end{minipage}
   \caption{Spectator-induced electromagnetic effect on directed flow of positive and negative pions in peripheral Pb+Pb collisions at $\sqrt{s_{NN}}=17.3$~GeV, shown at fixed values of pion transverse momentum: $p_T=75$~MeV/c (solid), $p_T=125$~MeV/c (dashed), $p_T=175$~MeV/c (dotted). The top and bottom panels correspond to different values of the pion emission time $t_E$.}
 \label{fig:pt}
%           }
\end{figure}
%%%%%%%%%%%%%%%%%%%%%%%%%%%%%%%%%%%%%%%%%%%%%%%%%%%%%%%%%%%%%%%%%%

Up to now, for simplicity, only the rapidity-dependence of $p_T$-integrated directed flow was discussed in this paper. In the present Section we study its dependence on the transverse momentum of the pion. This is shown in Fig.~\ref{fig:pt}. The electromagnetically-induced directed flow exhibits a strong dependence on pion $p_T$. In particular, this is valid close to beam rapidity where very large absolute values of $v_1$ (of the order of 0.6 for positive pions) are attained at the lowest considered value of transverse momentum, $p_T=75$~MeV/c. The absolute value of the electromagnetically-induced $v_1$ rapidly decreases with increasing $p_T$. On the other hand, little or no dependence on pion $p_T$ is apparent in the region of lower absolute values of $y/y_\mathrm{beam}$, down to midrapidity. The overall pattern of dependence on initial conditions, discussed in Section~\ref{initial} above, repeats itself also at fixed values of $p_T$. 

%\newpage
%.
%\newpage

%%%%%%%%%%%%%%%%%%%%%%%%%%%%%%%%%%%%%%%%%%%%%%%%%%%%%%%%%%%%%
\begin{figure}[t]             
%\begin{minipage}[t]{0.46\textwidth}
\centering
\includegraphics[width=0.62\textwidth]{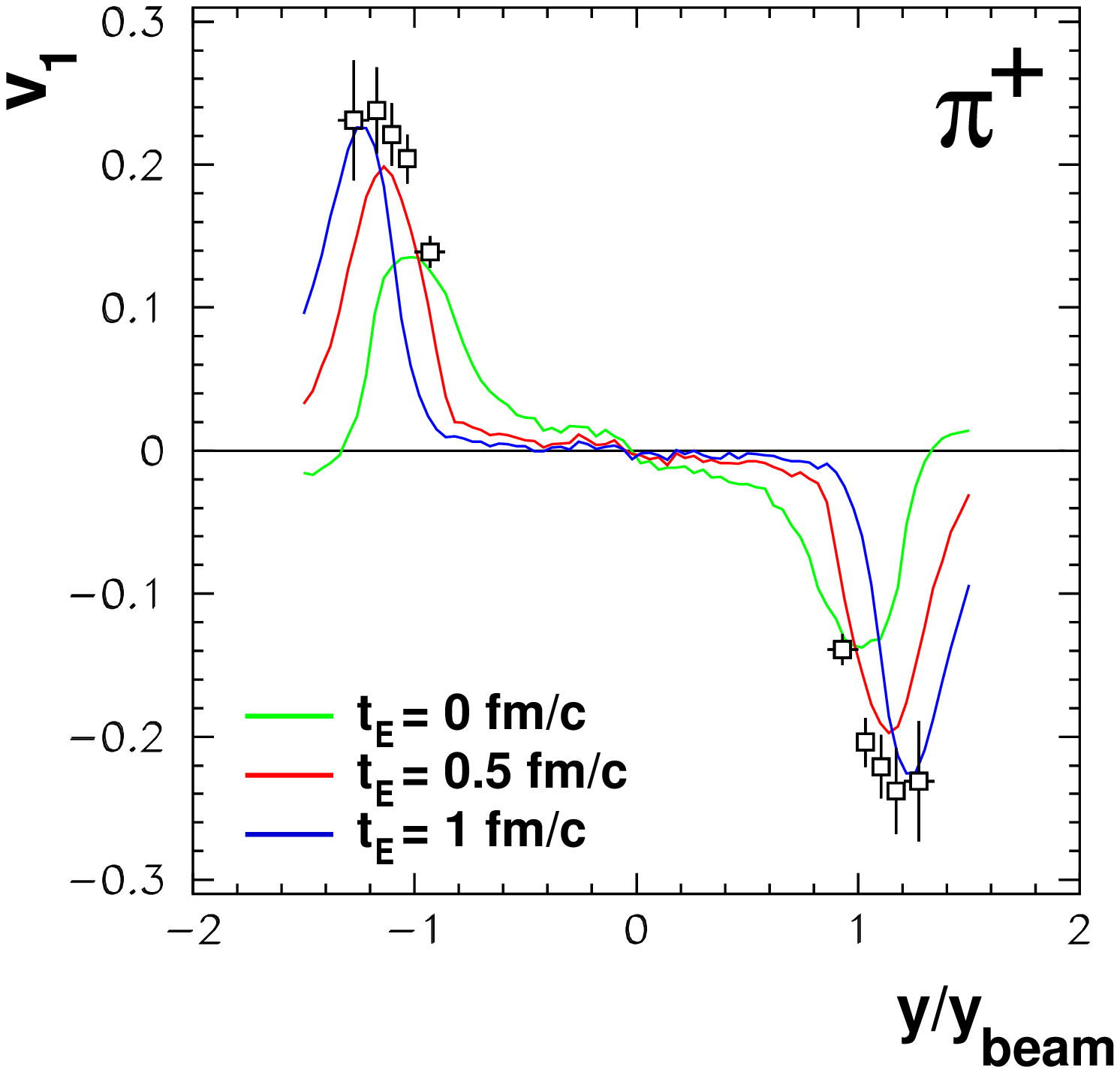}
%\end{minipage}
   \caption{Comparison between experimental data on directed flow of positive pions obtained by the WA98 experiment (the data points
 are redrawn
 from~\cite{wa98}), and our simulation of electromagnetically-induced directed flow of $\pi^+$. The three curves correspond to three different values of the assumed pion emission time $t_E$. Note: the data points at positive $y/y_\mathrm{beam}$ are
obtained by reflection about midrapidity.}
 \label{fig:comp}
%           }
\end{figure}
%%%%%%%%%%%%%%%%%%%%%%%%%%%%%%%%%%%%%%%%%%%%%%%%%%%%%%%%%%%%%%%%%%

%\newpage
%.
%\newpage

%-------------
\subsection{Comparison to experimental data}
\label{exp}
%-------------

Taking into account the charge-asymmetric nature of the electromagnetic effect discussed here, any comparison of our work to experimental data must involve 
 measurements of directed flow 
of
pions of {\em a given charge}, either $\pi^+$ or $\pi^-$. This constitutes an important practical difficulty as most of results published on the rapidity-dependence of $v_1$ involve (if particle identification is available at all) only ``charged pions'', i.e., summed $\pi^+$ and $\pi^-$. One data set which fulfills our requirement comes from the WA98 experiment, and contains a measurement of positive pions near target rapidity made by means of the Plastic Ball detector, in Pb+Pb collisions at $\sqrt{s_{NN}}=17.3$~GeV. This measurement is presented in~\cite{wa98}, but a more detailed description of the analysis can be found in~\cite{schlagheck}. 
The latter reference specifies the corresponding centrality definition as 
40-80\% of the Pb+Pb cross-section. Following the discussion made therein,
 we expect that differences w.r.t. our definition of the peripheral collision (Section~\ref{model}) would only have a 
small
effect on the actual values of $v_1$.

Fig.~\ref{fig:comp} shows the experimental data from WA98, superimposed with the rapidity-dependence of the electromagnetically-induced directed flow as obtained from our work. For the latter, three values of the pion emission time are assumed: $t_E=0$, $0.5$, and $1$~fm/c. Our results are integrated over $p_T$ from 0 to 1~GeV/c. The curves corresponding to $t_E=0$ and $1$~fm/c are the same as in Fig.~\ref{fig:initial}.

As it can be seen in the figure, the values of 
$v_1$ obtained from our simulation, and resulting exclusively from the spectator-induced electromagnetic interaction, appear comparable to those measured by 
experiment.

The above statement should clearly be taken with some caution. 
Account should be taken of the simplicity of our model, and of various detailed issues related to the measurement~\cite{wa98}. These 
problems
are beyond the scope of the present paper.

Nevertheless, it seems clear that a very sizeable part of the measured 
directed flow of $\pi^+$ {\em may come from the electromagnetic origin}. We note that this observation is consistent with our analysis of the influence of spectator charge on $\pi^+/\pi^-$ ratios in peripheral Pb+Pb collisions at the same energy. Here, pion emission times assumed in the range $t_E=0.5$-$1$~fm/c also 
result
in good agreement between our model and
 experimental data~\cite{ismd}.

%\newpage

%----------------------------
\section{Summary and Discussion}
\label{discussion}
%----------------------------

%In the present Section we shortly summarize the principal results of our work and discuss their most evident implications. 

Summing up the principal results of our work, the following can be said:

\begin{enumerate}
\item 
 We devised a simple model of the Pb+Pb collision (Section~\ref{model}). This model does not contain any initial azimuthal anisotropy in pion emission {but} includes the electromagnetic interaction between the pion charge and the two spectator systems. 
\item 
 This model predicts sizeable values of $v_1$ (electromagnetically-induced directed flow) for positive and negative pions in the final state of the collision. For positive pions, the predicted values are comparable to those seen in experimental data. As such, a very large part of positive pion directed flow may come from the electromagnetic interaction.
\item 
The sign of this effect is opposite for positively and negatively charged pions. This will result in a splitting of values of $v_1$ observed for $\pi^+$ and $\pi^-$.
\item 
The size of electromagnetically-induced directed flow, and therefore also the size of the electromagnetic splitting addressed above, depends on the pion emission time $t_E$. The latter parameter defines, in our model, the details of the space-time evolution of pion production, like e.g. the time of pion formation, or the position of the pion formation zone with respect to the spectator system(s).
\item 
 As it follows from the above, the electromagnetically-induced directed flow, and the resulting splitting of $v_1$ for $\pi^+$ and $\pi^-$, {\em bring new information on the space-time dynamics of the Pb+Pb reaction}. 

\end{enumerate}

This has, in our view, important consequences for future studies of directed flow. It is of course clear that our model contains important simplifications. Among others, it intentionally neglects the initial pion flow induced by the strong rather than the electromagnetic force\footnote{We keep in mind that the inclusion of the latter is necessary in order to describe the summed charged pion data (see, e.g.,~\cite{na49prc}).}. The inclusion of such phenomena is the domain of more detailed models and we leave it for future studies. However, it is clear from our work that the electromagnetic splitting of $v_1(\pi^+)$ and $v_1(\pi^-)$ addressed above will remain present also in more sophisticated phenomenological descriptions of the heavy ion collision. This inspires a set of proposals and remarks listed below.
\begin{enumerate}
\item[$\bullet$]
 The study of rapidity-dependence of directed flow made {\em separately} for pions of a given charge, i.e., $\pi^+$ and $\pi^-$, will bring additional information on the dynamics of the heavy ion collision. This information will add up to the already rich phenomenological content of summed charged pion flow studies. The electromagnetic splitting of $v_1$ for positive and negative pions will provide information about the position of the pion formation zone with respect to the spectator systems. Specifically, also the {\em lack} of such splitting at a given rapidity would imply that the formation
% (freeze-out)
 of pions occurs relatively far from the nearest spectator.
\item[$\bullet$]
 The situation should be similar for other produced particles, like e.g., charged kaons. The situation for final state protons should be at least partially different as protons at 
 %large 
 forward rapidity will come 
 also
 %mostly 
 from spectator fragmentation~\cite{Bartke09,na49spec,na49pc}. However, also in this case the role played by the electromagnetic repulsion of protons from the spectator systems should be investigated in detail.
\item[$\bullet$]
Last but not least, our analysis brings attention to {\em asymmetric} nuclear collisions, be it A+B, d+A, or p+A reactions. In such collisions, the electromagnetic influence of the two spectators would not cancel out at midrapidity, and therefore the role of single spectators could be isolated, yielding further information on the space-time evolution of the collision (see Fig.~\ref{fig:onespec} for comparison).
\end{enumerate}

All of the above observations are consistent with the results of our
study of spectator-induced electromagnetic forces acting on charged pion
ratios~\cite{twospec}. However, we feel that the precision of the
present experimental measurements of the size of directed flow, the key
role played by flow measurements in the present phenomenology of heavy
ion collisions, and the overall interest of the community in such
studies, make the consideration of the role played by electromagnetic interactions
important for future analyses. 

%----------------------------
\section{Data from the RHIC Beam Energy Scan}
\label{bes}
%----------------------------

In the context of the discussion made above, and based on the comparison between the results of our simulation and existing SPS experimental data, it is important to note that the RHIC Beam Energy Scan (BES) program can also address the search of the spectator-induced electromagnetic effects discussed here.

Specifically, preliminary data on directed flow of positive and negative pions from the STAR experiment have already been reported in~\cite{Pandit-identified-charged}.
The measured values of $v_1(\pi^+)$ and $v_1(\pi^-)$ indeed display a signal of splitting which increases with increasing rapidity. At positive rapidity, $v_1(\pi^+)$ remains below $v_1(\pi^-)$, which follows our predictions for the electromagnetically-induced directed flow,
formulated in Sections~\ref{onespec} and \ref{initial} above. 

Presenting final conclusions on the nature of this effect probably requires more precise modelling of the space-time mechanism of charged pion production, including also its centrality dependence.
 % as well as final experimental data. ??
However, we note that the overall magnitude of the splitting of $v_1$ values seen by experiment looks (roughly) comparable to that predicted in the preceding Sections, for the largest assumed values of the pion emission time $t_E$ and in the relatively limited range of $y/y_\mathrm{beam}$ available to the measurement~\cite{Pandit-identified-charged}.
{\em Thus, we consider that the observed splitting is indeed a signature of electromagnetically-induced directed flow}.

%----------------------------
\section{Conclusions}
\label{conclusions}
%----------------------------

The electromagnetic field induced by the presence of spectator systems
may result in sizeable distortions of the directed flow pattern observed
for pions of a given charge (that is, $\pi^+$ and $\pi^-$) produced in
heavy ion collisions. This implies the presence of the electromagnetic
splitting of $v_1$ for particles of opposite charges. The size of this
splitting depends on the space-time scenario of pion production and
therefore brings new information on the collision dynamics. The
comparison to experimental data suggests that a very large part of
directed flow observed for positive pions may come from 
the electromagnetic origin. 

New experimental and phenomenological studies are needed in order to
elucidate these questions, and possibly bring new insight into the
evolution of the heavy ion reaction in space and time. On the
experimental side, this implies measurements of particles of specific
charges, for symmetric and asymmetric nuclear reactions. On the
theoretical side, the possibility of verification of existing models of 
the heavy ion reaction by taking into account spectator-induced
electromagnetic effects should be investigated. Furthermore, studies of
the role of spectator-induced electromagnetic interactions for higher
harmonics of the azimuthal distribution would also be highly indicated.\\

{\bf Acknowledgments}\\

This work was supported by the Polish National Science Centre 
(on the basis of decision no. DEC-2011/03/B/ST2/02634).

%\vspace*{1cm}
%[bozek] ???????????


\begin{thebibliography}{999}

%%%%%%%%%%%%%%%%%%%%%%%%%%%%
\bibitem{Snellings1999}
  R.~J.~M.~Snellings, H.~Sorge, S.~A.~Voloshin, F.~Q.~Wang, N.~Xu,
  %``Novel rapidity dependence of directed flow in high-energy heavy ion collisions,''
  Phys.\ Rev.\ Lett.\  {\bf 84}, 2803 (2000).
  %[nucl-ex/9908001].
  %%CITATION = NUCL-EX/9908001;%%
  %59 citations counted in INSPIRE as of 25 Mar 2013

\bibitem{Csernai2012}
  L.~P.~Csernai, G.~Eyyubova, and V.~K.~Magas,
  %``New method for measuring longitudinal fuctuations and directed flow in ultrarelativistic heavy ion reactions,''
  Phys.\ Rev.\ C {\bf 86}, 024912 (2012).
 %  [arXiv:1204.5885 [hep-ph]].
  %%CITATION = ARXIV:1204.5885;%%
  %1 citations counted in INSPIRE as of 25 Mar 2013

\bibitem{Luzum2010}
  M.~Luzum and J.-Y.~Ollitrault,
  %``Directed flow at midrapidity in heavy-ion collisions,''
  Phys.\ Rev.\ Lett.\  {\bf 106}, 102301 (2011).
 % [arXiv:1011.6361 [nucl-ex]].
  %%CITATION = ARXIV:1011.6361;%%
  %24 citations counted in INSPIRE as of 25 Mar 2013

\bibitem{Retinskaya2012}
  E.~Retinskaya, M.~Luzum, and J.-Y.~Ollitrault,
  %``Directed flow at midrapidity in $\sqrt{s_{NN}}=2.76$ TeV Pb+Pb collisions,''
  Phys.\ Rev.\ Lett.\  {\bf 108}, 252302 (2012).
 % [arXiv:1203.0931 [nucl-th]].
  %%CITATION = ARXIV:1203.0931;%%
  %17 citations counted in INSPIRE as of 25 Mar 2013

\bibitem{florkowski}   %
W.~Florkowski,
  {\em Phenomenology of Ultra-Relativistic Heavy-Ion Collisions,}
  Singapore, World Scientific (2010).
%%%%%%%%%%%%%%%%%%%%%%%

\bibitem{bozek}
P.~Bo\.zek,
  %``Flow and interferometry in 3+1 dimensional viscous hydrodynamics,''
  Phys.\ Rev.\ C {\bf 85}, 034901 (2012),\\
%  [arXiv:1110.6742 [nucl-th]].
  %%CITATION = ARXIV:1110.6742;%%
  %25 citations counted in INSPIRE as of 25 Mar 2013
  %
  P.~Bo\.zek and I.~Wyskiel,
  %``Directed flow in ultrarelativistic heavy-ion collisions,''
  Phys.\ Rev.\ C {\bf 81}, 054902 (2010).
 % [arXiv:1002.4999 [nucl-th]].
  %%CITATION = ARXIV:1002.4999;%%
  %25 citations counted in INSPIRE as of 25 Mar 2013

\bibitem{poskanzerprc}
  A.~M.~Poskanzer and S.~A.~Voloshin,
  %``Methods for analyzing anisotropic flow in relativistic nuclear collisions,''
  Phys.\ Rev.\ C {\bf 58}, 1671 (1998).
 % [nucl-ex/9805001].

%%%%%%%%%%%%%%%%%%%%%%%%%%%%%%%%%%%%%
\bibitem{Andronic2001}
  A.~Andronic {\it et al.}  (FOPI Collaboration),
  %``Differential directed flow in Au+Au collisions,''
  Phys.\ Rev.\ C {\bf 64}, 041604 (2001).
 % [nucl-ex/0108014].
  %%CITATION = NUCL-EX/0108014;%%
  %28 citations counted in INSPIRE as of 25 Mar 2013

\bibitem{Barrette2000}
  J.~Barrette {\it et al.}  (E877 Collaboration),
  %``Directed flow of anti-protons in Au + Au collisions at AGS,''
  Phys.\ Lett.\ B {\bf 485}, 319 (2000).
  %[nucl-ex/0004002].
  %%CITATION = NUCL-EX/0004002;%%
  %6 citations counted in INSPIRE as of 25 Mar 2013

\bibitem{Barrette1998}
  J.~Barrette {\it et al.}  (E877 Collaboration),
  %``Directed flow of light nuclei in Au + Au collisions at AGS energies,''
  Phys.\ Rev.\ C {\bf 59}, 884 (1999).
  %[nucl-ex/9805006].
  %%CITATION = NUCL-EX/9805006;%%
  %17 citations counted in INSPIRE as of 25 Mar 2013

\bibitem{Adams2005}
  J.~Adams {\it et al.}  (STAR Collaboration),
  %``Directed flow in Au+Au collisions at s(NN)**(1/2) = 62-GeV,''
  Phys.\ Rev.\ C {\bf 73}, 034903 (2006).
  %[nucl-ex/0510053].
  %%CITATION = NUCL-EX/0510053;%%
  %40 citations counted in INSPIRE as of 25 Mar 2013

\bibitem{na49prc}
  C.~Alt {\it et al.} (NA49 Collaboration),
  %``Directed and elliptic flow of charged pions and protons in Pb + Pb collisions at 40-A-GeV and 158-A-GeV,''
  Phys.\ Rev.\ C {\bf 68}, 034903 (2003).
  %[nucl-ex/0303001].
  %%CITATION = NUCL-EX/0303001;%%
  %244 citations counted in INSPIRE as of 20 Mar 2013

\bibitem{Back2005}
  B.~B.~Back {\it et al.}  (PHOBOS Collaboration),
  %``Energy dependence of directed flow over a wide range of pseudorapidity in Au + Au collisions at RHIC,''
  Phys.\ Rev.\ Lett.\  {\bf 97}, 012301 (2006).
 %[nucl-ex/0511045].
  %%CITATION = NUCL-EX/0511045;%%
  %53 citations counted in INSPIRE as of 25 Mar 2013

\bibitem{Abelev2008}
  B.~I.~Abelev {\it et al.}  (STAR Collaboration),
  %``System-size independence of directed flow at the Relativistic Heavy-Ion Collider,''
  Phys.\ Rev.\ Lett.\  {\bf 101}, 252301 (2008).
 %[arXiv:0807.1518 [nucl-ex]].
  %%CITATION = ARXIV:0807.1518;%%
  %55 citations counted in INSPIRE as of 25 Mar 2013

\bibitem{Adamczyk2011}
  L.~Adamczyk {\it et al.}  (STAR Collaboration),
  %``Directed Flow of Identified Particles in Au + Au Collisions at $\sqrt{s{_NN}} = 200$ GeV at RHIC,''
  Phys.\ Rev.\ Lett.\  {\bf 108}, 202301 (2012).
 %[arXiv:1112.3930 [nucl-ex]].
  %%CITATION = ARXIV:1112.3930;%%
  %6 citations counted in INSPIRE as of 25 Mar 2013

\bibitem{Selyuzhenkov2011}
  I.~Selyuzhenkov,
  %``Charged particle directed flow in Pb-Pb collisions at $\sqrt{s_{NN}}=2.76$ TeV measured with ALICE at the LHC,''
  J.\ Phys.\ G {\bf 38}, 124167 (2011).
 %[arXiv:1106.5425 [nucl-ex]].
  %%CITATION = ARXIV:1106.5425;%%
  %10 citations counted in INSPIRE as of 25 Mar 2013

\bibitem{wa98}   %
  H.~Schlagheck (WA98 Collaboration),
  %``Thermalization and flow in 158-GeV/A Pb + Pb collisions,''
  Nucl.\ Phys.\ A {\bf 663}, 725 (2000).
  %[nucl-ex/9909005].
  %%CITATION = NUCL-EX/9909005;%%
  %3 citations counted in INSPIRE as of 20 Mar 2013

%%%%%%%%%%%%%%%%%%%%%%%%%%%%%%%%%%%%%%

\bibitem{twospec}   %2
 A. Rybicki, A. Szczurek, Phys. Rev. C {\bf 75}, 054903 (2007).

\bibitem{rybickizako}
  A.~Rybicki,
  %``What Is the role of nuclear effects in ultrarelativistic reactions at 158-GeV/nucleon?,''
  Acta Phys.\ Polon.\ B {\bf 42}, 867 (2011), and references therein. \\
  See also: G. Ambrosini {\it et al.} (NA52 Collaboration), New Jour.~Phys. {\bf 1}, 23 (1999).

%Ambrosini99a?

%%%%%%%%%%%%%%%%here refs about precedent works
\bibitem{Bartke09}
  J.~Bartke,
  {\em Introduction to Relativistic Heavy Ion Physics,}
%\href{http://www.slac.stanford.edu/spires/find/hep/www?irn=8472912}{SPIRES entry}
%{\it  Hackensack, USA: World Scientific (2009)}
% 224 p}
{  Hackensack, USA: World Scientific (2009)}, and references therein.

\bibitem{Yagoda52}
H.~Yagoda, Phys. Rev. {\bf 85}, 891 (1952).

\bibitem{Friedlander62}
E.~M.~Friedl\"ander, Phys. Lett. {\bf 2}, 38 (1962).

\bibitem{Beneson79}
W. Beneson {\it et al.}, Phys. Rev. Lett. {\bf 43}, 683 (1979).

\bibitem{Sullivan82}
J.~P.~Sullivan {\it et al.}, Phys. Rev. C {\bf 25}, 1499 (1982).

\bibitem{Karnaukhov06}
V.~A.~Karnaukhov {\it et al.}, Phys.\ Atom.\ Nucl.\  {\bf 69}, 1142 (2006).

\bibitem{Ayala97}
A.~Ayala, J.~Kapusta, Phys. Rev. C {\bf 56}, 407 (1997).

\bibitem{Ayala99}
A.~Ayala, S.~Jeon, J.~Kapusta, Phys. Rev. C {\bf 59}, 3324 (1999).

\bibitem{Ahle98}
L.~Ahle {\it et al.} (E802 Collaboration), Phys.~Rev.\ C {\bf 57}, R466 (1998).


\bibitem{Lacasse96}
R.~Lacasse {\it et al.} (E877 Collaboration), Nucl.~Phys.~A~{\bf 610}, 153c (1996).

\bibitem{Xu96}
N.~Xu {\it et al.} (NA44 Collaboration),
 Nucl.~Phys.~A {\bf 610}, 175c (1996).

\bibitem{Libbrecht79}
K. G. Libbrecht and S.~E. Koonin, Phys. Rev. Lett. {\bf 43}, 1581 (1979).

\bibitem{Cugnon81}
J. Cugnon and S.~E. Koonin, Nucl. Phys. A {\bf 355}, 477 (1981).

\bibitem{Gyulassy81}
M.~Gyulassy and S.~K.~Kauffmann, Nucl. Phys. A {\bf 362}, 503 (1981).

\bibitem{Bonasera87}
A. Bonasera and G. F. Bertsch, Phys. Lett. B {\bf 195}, 521 (1987).

\bibitem{Li95}
B.-A. Li, Phys. Lett. B {\bf 346}, 5 (1995).

\bibitem{Barz98}
H.~W. Barz, J.~P. Bondorf, J.~J. Gaardhoje and H.~Heiselberg,
Phys. Rev. C {\bf 57}, 2536 (1998).

\bibitem{Osada96}
  T.~Osada, S.~Sano, M.~Biyajima and G.~Wilk,
  %``What information can we obtain from the yield ratio pi-/pi+ in heavy  ion
  %collisions?,''
  Phys.\ Rev.\  C {\bf 54}, 2167 (1996).
  %[arXiv:hep-ph/9606366].
  %%CITATION = PHRVA,C54,2167;%%

%%%%%
\bibitem{mizutori}
  S.~Mizutori, J.~Dobaczewski, G.~A.~Lalazissis, W.~Nazarewicz and P.~G.~Reinhard,
  %``Nuclear skins and halos in the mean field theory,''
  Phys.\ Rev.\ C {\bf 61}, 044326 (2000),\\
%  [nucl-th/9911062].
  %%CITATION = NUCL-TH/9911062;%%
  %65 citations counted in INSPIRE as of 22 Mar 2013
A. Trzci\'nska, Ph. D. dissertation, Heavy Ion Laboratory, Warsaw
University, 2001.

\bibitem{pp}
C.~Alt {\it et al.} (NA49 Collaboration), 
%   ``Inclusive production of charged pions in p p collisions at 158-GeV/c beam
%momentum,''
  {Eur.\ Phys.\ J.\ C} {\bf 45}, 343 (2006).
%   numerical data and interpolation: http://spshadrons.web.cern.ch

\bibitem{wnm}
A.~Bia\l{}as, M.~Bleszy\'nski and W.~Czy\.z,
%``Multiplicity Distributions In Nucleus-Nucleus Collisions At High-Energies,''
Nucl.\ Phys.\ B {\bf 111}, 461 (1976).

%%%%%%%%%%%%%%%%
\bibitem{star-Pandit}
  Y.~Pandit (STAR Collaboration),
  %``Beam Energy Dependence of Directed and Elliptic Flow Measurement from the STAR Experiment,''
  J.\ Phys.\ Conf.\ Ser.\  {\bf 316}, 012001 (2011).
%  [arXiv:1109.2799 [nucl-ex]].
  %%CITATION = ARXIV:1109.2799;%%
  %4 citations counted in INSPIRE as of 22 Mar 2013

\bibitem{schlagheck}
 H.~Schlagheck, Ph.~D. dissertation, University of M\"unster, 1998.

\bibitem{ismd}
M.~K\l{}usek-Gawenda, 
E.~Kozik,
A.~Rybicki, 
I.~Sputowska, 
A.~Szczurek,
  %``Strong and Electromagnetic Forces in Heavy Ion Collisions,''
  arXiv:1303.6423 [nucl-ex].
  %%CITATION = ARXIV:1303.6423;%%
 % {\em Strong and Electromagnetic Forces in Heavy Ion Collisions}, in: proc. Int. Symposium on Multiparticle Dynamics - ISMD2012, Kielce, 16 Sept. 2012, to appear in Acta Phys.\ Polon.\ B
 %[arXiv:xxxx.xxxx [nucl-th]].

\bibitem{na49spec}
  H.~Appelsh\"auser {\it et al.} (NA49 Collaboration),
  %``Spectator nucleons in Pb + Pb collisions at 158-A-GeV,''
  Eur.\ Phys.\ J.\ A {\bf 2}, 383 (1998).
  %%CITATION = EPHJA,A2,383;%%
  %29 citations counted in INSPIRE as of 25 Mar 2013

\bibitem{na49pc}
  B.~Baatar {\it et al.} (NA49 Collaboration),
  %``Inclusive production of protons, anti-protons, neutrons, deuterons and tritons in p+C collisions at 158 GeV/c beam momentum,''
  arXiv:1207.6520 [hep-ex].
  %%CITATION = ARXIV:1207.6520;%%
  %1 citations counted in INSPIRE as of 25 Mar 2013

\bibitem{Pandit-identified-charged}
  Y.~Pandit (STAR Collaboration),
  %``Directed flow of Identified Charged Particles from the RHIC Beam Energy Scan,''
  Acta Phys.\ Polon.\ Supp.\  {\bf 5}, 439 (2012).
%  [arXiv:1112.0842 [nucl-ex]].
  %%CITATION = ARXIV:1112.0842;%%
  %1 citations counted in INSPIRE as of 25 Mar 2013



\end{thebibliography}
\end{document}